\documentclass[aps,prd,12pt,nofootinbib,superscriptaddress]{revtex4-2}
\usepackage{bm}
\usepackage{ulem}
\usepackage{comment}
\usepackage{amsmath}
\usepackage{amssymb}
\usepackage{booktabs}
\usepackage{graphicx}
\usepackage{float}
\usepackage{orcidlink}
\usepackage{comment}
\usepackage{hyperref}
\usepackage{CJK}
\usepackage{color}
\hypersetup{
	colorlinks=true,
	linkcolor=red,
	citecolor=blue,
}

\begin{document}
\begin{CJK*}{UTF8}{gbsn}
\title{Distinguishing dark matter halos with Extreme mass ratio inspirals}

\author{Yang Zhao (赵阳)
\orcidlink{0009-0003-7436-8668}}
\email{zhaoyangedu@hust.edu.cn}
\affiliation{School of Physics, Huazhong University of Science and Technology, 1037 LuoYu Rd, Wuhan, Hubei 430074, China}

\author{Ning Dai (戴宁) \orcidlink{0000-0002-0867-6764}}
\email{daining@hust.edu.cn}
\affiliation{School of Physics, Huazhong University of Science and Technology, 1037 LuoYu Rd, Wuhan, Hubei 430074, China}

\author{Yungui Gong \orcidlink{0000-0001-5065-2259}}
\email{Corresponding author. yggong@hust.edu.cn}
\affiliation{Institute of Fundamental Physics and Quantum Technology, Department of Physics, School of Physical Science and Technology, Ningbo University, Ningbo, Zhejiang 315211, China}
\affiliation{School of Physics, Huazhong University of Science and Technology, 1037 LuoYu Rd, Wuhan, Hubei 430074, China}


\begin{abstract}
Using the static, spherically symmetric metric for a black hole (BH) immersed in dark matter (DM) halo characterized by Hernquist, Burkert, and Navarro-Frenk-White (NFW) density distributions, 
we calculate the orbital periods and precessions, along with the evolution of the semi-latus rectum and eccentricity for extreme mass ratio inspirals (EMRIs) surrounded by DM halos. 
For the Hernquist model, we find that the gravitational force exerted by the central BH is decreased by DM halos, while DM halos put additional gravitational force on the SCO.
The presence of both Burkert-type and NFW-type DM halos enhances the gravitational force acting on the SCO, 
resulting in a decrease in the period $P$, 
with the decrease depending on $M/a_0^2$;
additionally, we find that the reduction in orbital precession due to DM halos is influenced by $M/a_0^2$.
The presence of DM halos leads to a slower evolution of EMRIs within Hernquist-type halos,
while it accelerates evolution for EMRIs in Burkert-type and NFW-type halos;
furthermore, it slows the decrease of eccentricity across all three types of DM halos.
By calculating the number of orbital cycles and the gravitational waveform mismatches among these three types of DM halos, as well as between scenarios with and without DM halos, 
we find that DM halos can be detected when $M/a_0>10^{-5}$, $M/a_0>10^{-3}$, 
and $M/a_0>10^{-3}$ for Hernquist-type, NFW-type, and Burkert-type DM halos, respectively.  
Additionally, we can distinguish between NFW-type and Burkert-type DM halos when $M/a_0> 10^{-3}$; 
NFW-type and Hernquist-type DM halos, as well as Burkert-type and Hernquist-type DM halos, can be distinguished when $M/a_0> 10^{-5}$.

\end{abstract}

\maketitle
\end{CJK*}

\section{Introduction}

Although there is a large amount of observational evidence from different scales supporting the existence of dark matter (DM), which accounts for 26\% of the total mass of the Universe \cite{Zwicky:1933gu,Rubin:1970zza, Rubin:1980zd,Bertone:2004pz,Clowe:2006eq,WMAP:2010qai,Bertone:2016nfn,Planck:2018vyg}, we still know little about its nature and origin. 
The study of DM is crucial for understanding the formation and evolution of the Universe, as well as for potential breakthroughs in fundamental physics \cite{Bertone:2004pz,Bertone:2016nfn}. 
DM may cluster around galaxies and form DM halos \cite{Navarro:1996gj, Gondolo:1999ef, Burkert:1995yz, Taylor:2002zd, Ullio:2001fb, Moore:1997sg, Feng:2010gw}. 
Depending on the size, mass, and structure of a galaxy, various distributions of DM halos are favored \cite{Taylor:2002zd}.
For example, the DM halo around a dwarf galaxy, composed of several billion stars, can be described by the Burkert distribution \cite{Burkert:1995yz}. 
The Hernquist model is suited for describing the profiles observed in bulges and elliptical galaxies  \cite{Hernquist:1990be}. 
For galaxies with a high DM content, the Navarro-Frenk-White (NFW) model is predominantly used \cite{Navarro:1994hi}. 
The Taylor-Silk model is applied to describe the lightest neutralino, which is a candidate for cold DM in the Universe \cite{Taylor:2002zd}.

DM halos have been indirectly observed mainly through rotation curves \cite{Rubin:1970zza, Rubin:1980zd} and large-scale structures \cite{Allen:2002eu}. 
However, these methods are unable to accurately determine the nature of DM halos in regions with strong gravitational fields \cite{Sadeghian:2013laa}.
The detection of gravitational waves (GWs) offers new opportunities for studying DM halos \cite{LIGOScientific:2016aoc,LIGOScientific:2016emj,LIGOScientific:2018mvr,LIGOScientific:2020ibl,LIGOScientific:2021usb,KAGRA:2021vkt,Eda:2013gg}. 

A stellar-mass compact object (SCO) inspiraling into a supermassive black hole (SMBH) forms a binary system known as extreme mass ratio inspirals (EMRIs) \cite{Babak:2017tow}. 
The SMBH, with a mass ranging from $10^6~M_{\odot}$ to $10^{10}~M_{\odot}$, 
is believed to exist at the center of a galaxy \cite{Kormendy:2013dxa} and is often surrounded by DM halos. 
The presence of DM halos affects the motion of EMRIs within these halos. 
The rich information about the spacetime geometry surrounding the SMBH is encoded in the GWs, making EMRIs an excellent source for studying the nature of DM halos.
EMRIs emit millihertz GWs that are expected to be observed by future space-based GW detectors such as the Laser Interferometer Space Antenna (LISA) \cite{Danzmann:1997hm,LISA:2017pwj,Colpi:2024xhw}, Taiji \cite{Hu:2017mde} and TianQin \cite{TianQin:2015yph,Gong:2021gvw}. 
Therefore, EMRIs are anticipated to provide valuable constraints on DM halos \cite{Yunes:2011ws, Kocsis:2011dr, Eda:2014kra, Barack:2018yly, Cardoso:2019rou, Hannuksela:2018izj}.

There are a lot of studies on EMRIs within DM halos.
The effects of DM halos were usually modeled at the Newtonian level \cite{Eda:2013gg, Kavanagh:2020cfn, Dai:2021olt}. 
In Ref. \cite{Cardoso:2021wlq}, the spacetime geometry of a spherically symmetric, static, non-vacuum black hole (BH) generated by a DM distribution with the Hernquist density profile was investigated. 
Exact solutions for a SMBH surrounded by various types of DM halos were derived in \cite{Konoplya:2022hbl, Jusufi:2022jxu,Shen:2023erj, Figueiredo:2023gas, Speeney:2024mas}. 
The evolution of EMRIs, influenced by dynamical friction and accretion, affects both orbital motion and GW emission.
The influences of dynamical friction and accretion result in variations in the phase of the GW waveform \cite{Eda:2013gg, Barausse:2014tra,Antonini:2016gqe,Dai:2023cft,Zhang:2024ugv}.
Numerical methods for calculating orbital geodesics were proposed in \cite{Destounis:2022obl}.
Using the BH perturbation method, a generic formalism for calculating GW fluxes from EMRIs within DM halos was developed in \cite{Cardoso:2022whc,Figueiredo:2023gas,Speeney:2024mas}. 

The density distribution of a DM halo can reveal its intrinsic properties, 
motivating us to investigate whether different types of DM halos can be distinguished using GWs from EMRIs. 
In \cite{Cole:2022yzw}, the distinction between accretion disks, dark matter spikes and clouds of ultra-light scalar fields was studied.
In this paper, we study the eccentric orbital motions and GW emissions of EMRIs in galaxies surrounded by various types of DM halos. 
Considering the effects of GW radiation, DM accretion and dynamical friction, we compare the orbital evolution and waveforms across different types of DM halos.

The paper is organized as follows: 
In Sec. \ref{motion}, we introduce the method for studying GW emission, accretion and dynamical friction by EMRIs in spherically symmetric, non-vacuum BH spacetime with three different galaxy models. 
We calculate the energy and angular momentum fluxes in the background of a SMBH surrounded by different types of DM halos and analyze the combined effects on orbital evolution influenced by GW radiation, accretion and dynamical friction. 
In Sec. \ref{waveform}, we use the "Numerical Kludge" method \cite{Babak:2006uv,Gair:2005ih,Chua:2017ujo} to calculate GWs from eccentric EMRIs in galaxies with DM halos. 
We compute the orbital cycles and the mismatch in GW waveforms from EMRIs, comparing scenarios with and without DM halos, as well as comparing among three different types of DM halos.
Sec. \ref{conclusion} is devoted to conclusions and discussions. 
We use units where $G=c=1$.

\section{METHOD}
\label{motion}

For a general model of galaxy, the density of DM halos can be described by \cite{Taylor:2002zd, Graham:2005xx}
\begin{equation}
\label{rho-1}  \rho(r)=2^{(\gamma-\alpha)/k}\rho_0(r/a_0)^{-\alpha}(1+r^k/a_0^k)^{-(\gamma-\alpha)/k},
\end{equation}
where $a_0$ represents the typical length scale of a galaxy, $\rho_0$ denotes the halo density at $r=a_0$, and $\alpha$, $\beta$, and $\gamma$ are parameters that depend on the nature of DM halos and galaxies. 
The Burkert model ($\alpha=1, \gamma=3, k=2$) is primarily used to characterize the dwarf galaxies \cite{Burkert:1995yz, Salucci:2000ps}.
For galaxies with a significant amount of DM, the NFW model ($\alpha=1, \gamma=3, k=1$) is appropriate \cite{Navarro:1994hi}.
The Hernquist model ($\alpha=1, \gamma=4, k=1$) effectively describes the profiles observed in the bulges and elliptical galaxies \cite{Hernquist:1990be}. 

The energy-momentum tensor of galaxy harboring a SMBH with mass $M_\text{BH}$ is assumed to be represented by the anisotropic fluid \cite{Cardoso:2021wlq, Benson:2010de}
\begin{equation}
    T^\mu_\nu={\rm diag}(-\rho_\text{DM},0,P_\text{t},P_\text{t}),
\end{equation}
where $\rho_\text{DM}$ is the density distribution of the DM halo surrounding the SMBH and $P_\text{t}$ denotes the tangential pressure.
Following \cite{Figueiredo:2023gas, Cardoso:2021wlq}, we choose the density profiles as
\begin{equation}
\label{rho-2}
    \rho_\text{DM}(r)=\rho(r)\left(1-\frac{2M_\text{BH}}{r}\right).
\end{equation}
The spacetime surrounding the SMBH is described by a static and spherically symmetric metric \cite{Cardoso:2021wlq}
\begin{equation}
\label{b_metric}
    ds^2=-A(r) dt^2+B(r) dr^2+r^2(d\theta^2+\sin^2\theta\,d\phi^2),
\end{equation}
where $B(r)=\left[1-2m(r)/r\right]^{-1}$ and $m(r)=4\pi r^2 \rho_\text{DM}(r)$ is the mass function.
Using Einstein equation, we obtain
\begin{equation}
\begin{split}
\label{metric1}
    \frac{A'(r)}{A(r)}&=\frac{2m(r)/r}{r-2m(r)},\\
    P_t(r)&=\frac{m(r)/2}{r-2m(r)}\rho_\text{DM}(r),
\end{split}
\end{equation}
where the prime denotes the differentiation with respect to $r$.
The metric can be solved using the following boundary conditions \cite{Figueiredo:2023gas}
\begin{equation}
\begin{split}
\label{metric2}          
    m(R_\text{s})&=M_{\text{BH}},\\
    m(r\rightarrow r_\text{out})&=M+M_{\text{BH}},\\
    A(r\rightarrow r_\text{out})&=1-\frac{2(M_{\text{BH}}+M)}{r},
\end{split}
\end{equation}
where $R_\text{s}=2M_\text{BH}$ is the event horizon of a black hole, $r_\text{out}$ corresponds to spatial infinity and $M$ is the mass of DM halos.
For the NFW and Burkert density distribution, $m(r)$ diverges logarithmically with $r$.
We assume that the density of the halos vanishes outside the galaxy and the radius of the galaxy is $r=r_\text{c}$, such that $M(r>r_\text{c})=0$ \cite{Navarro:1996gj}. 
Choosing $r_\text{out}=10^6 a_0>r_\text{c}=5a_0$, which corresponds to our numerical approximation of spatial infinity \cite{Figueiredo:2023gas, Konoplya:2022hbl}, we numerically solve the metric with Eqs. \eqref{metric1} and \eqref{metric2}.

\begin{figure*}[htp]
    \centering{$
    \begin{array}{cc}
    \includegraphics[width=0.45\textwidth]{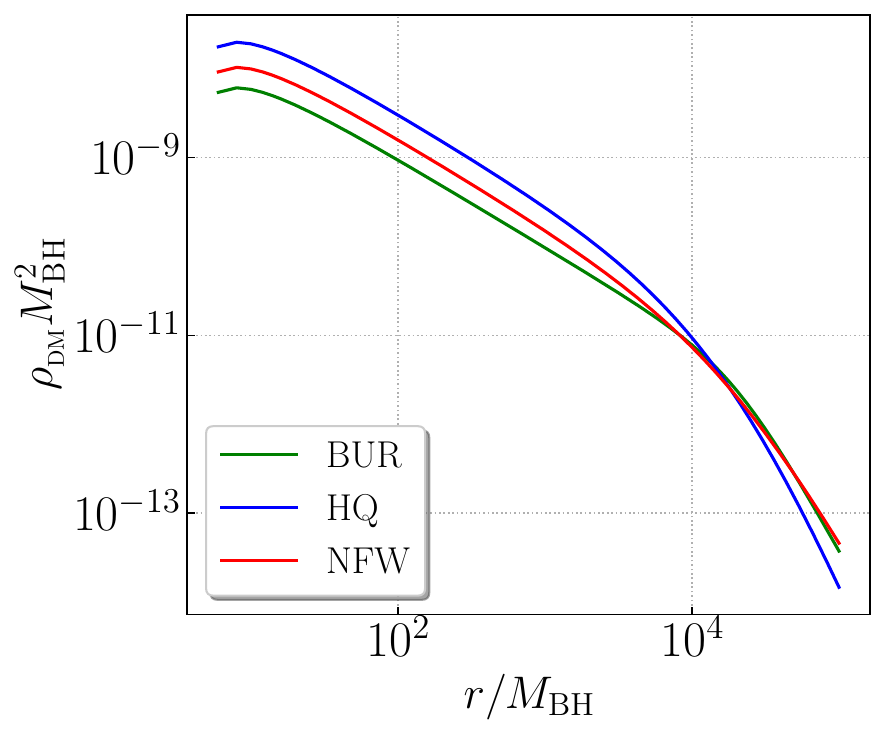}\quad \quad &
    \includegraphics[width=0.45\textwidth]{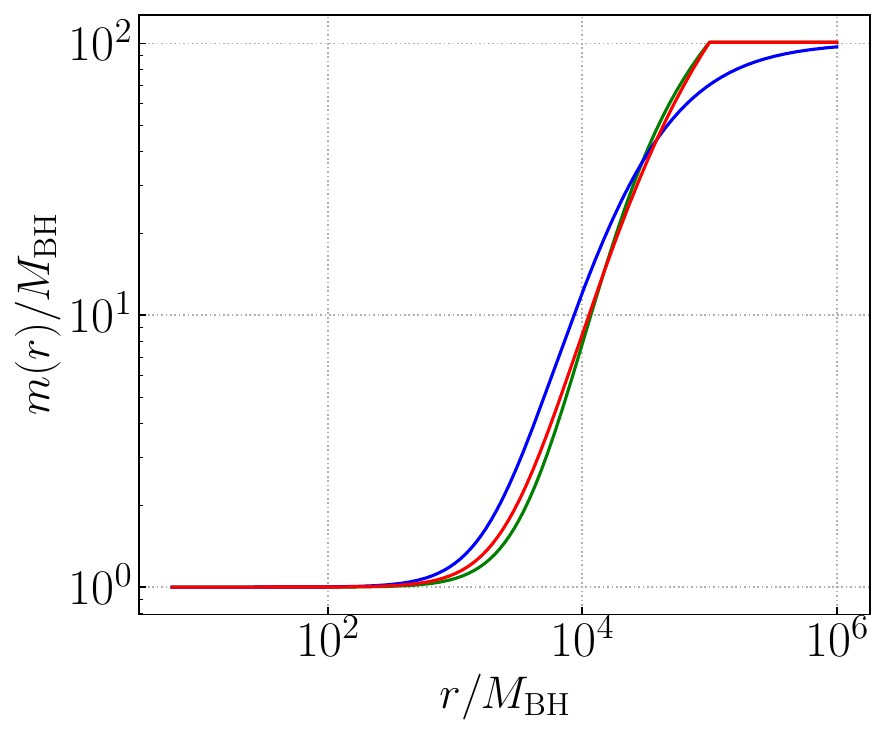}
    \end{array}$}
    \caption{
    The density distributios of the DM halos and the mass function,
    ``BUR", ``HQ" and ``NFW" represent the results calculated with Burkert-type, Hernquist-type, and NFW-type DM halos, respectively.
    The mass of central SMBH is taken as $M_\text{BH}=10^6M_\odot$,
    the length scale and the mass of DM halos are chosen as $a_0=100M$ and $M=100M_\text{BH}$, respectively.
    }
\label{fig:1}
\end{figure*}

The halo density $\rho(r)$ and mass functions $m(r)$ are shown in Fig. \ref{fig:1}. 
From the left panel of Fig. \ref{fig:1}, we see that for the range $6M_\text{BH}<r<1000M_\text{BH}$, 
the density distribution in the Hernquist model is the highest, while in the Burkert model, it is the lowest.
When $r>20000M_\text{BH}$, the Hernquist-type density distribution becomes the smallest.
From the right panel of Fig. \ref{fig:1}, we observe that within the range $6M_\text{BH}<r<10000M_\text{BH}$, the mass of the DM halos is greatest in the Hernquist model and smallest in the Burkert model.
When $r>10^5M_\text{BH}$, the Hernquist-type density distribution again becomes the smallest. For other parameter choices of $a_0$ and $M$, the results are similar.
From the approximate analytical expressions presented in the appendix \ref{APPENDIX1}, 
we see that the density is $\rho(r)\sim M/(r a_0^2)$,
resulting a constant force $\sim M/a_0^2$ acting on the SCO.

For a SCO inspiraling into a SMBH surrounded by DM halos,
the geodesic motion of the SCO can be parameterized by the semi-latus rectum $p$ and eccentricity $e$
\begin{equation}
\label{rpe}
r=\frac{p}{1+e \cos\chi},
\end{equation}
where $\chi$ is the orbital parameter,
$\chi$ varies monotonically from $\chi=2k\pi$ at the periastron $r_1=p/(1+e)$, to $\chi=(2k+1)\pi$ at the apastron $r_2=p/(1-e)$.
 
From Eq. \eqref{b_metric}, we define two conserved quantities
\begin{equation}
\begin{split}
\label{E-L1}
    E/\mu&=\epsilon=-u_0=\frac{\sqrt{r_2^2-r_1^2}\sqrt{A(r_1)A(r_2)}}{r_2^2A(r_1)-r_1^2A(r_2)},\\
    L/\mu&=h=u_\phi=\frac{r_1 r_2\sqrt{A(r_2)-A(r_1)}}{r_2^2A(r_1)-r_1^2A(r_2)},
\end{split}
\end{equation}
where $u_\alpha=g_{\alpha\beta}dx^\beta/d\tau$ is the four velocity,  
$E$ and $L$ are the orbital energy and angular momentum, $\mu$ is the mass of the SCO.
Considering the motion of the SCO in the equatorial plane $(\theta=\frac{1}{2}\pi)$ of the central SMBH spacetime,
we write the equation of motion as
\begin{equation}
\begin{split}
\label{radial}
    1+\left(\frac{dr}{d\tau}\right)^2\left(1-\frac{2m(r)}{r}\right)^{-1}+\frac{h^2}{r^2}=\frac{\epsilon}{A(r)}.
\end{split}
\end{equation}
Combining Eqs. \eqref{rpe}, \eqref{E-L1}, and \eqref{radial}, we get
\begin{equation}
\begin{split}
\label{orbital3}
    \frac{dt}{d\chi}&=\frac{\epsilon}{A(r)}\frac{dr}{d\chi}\left\{\left(1-\frac{2m(r)}{r}\right)\left[-1+\frac{\epsilon^2}{A(r)}-\frac{h^2}{r^2}\right]\right\}^{-1/2},\\
\end{split}
\end{equation}
\begin{equation}
\begin{split}
    \label{orbital2}
    \frac{d\phi}{d\chi}&=\frac{h}{r^2}\frac{dr}{d\chi}\left\{\left(1-\frac{2m(r)}{r}\right)\left[-1+\frac{\epsilon^2}{A(r)}-\frac{h^2}{r^2}\right]\right\}^{-1/2}.\\
\end{split}
\end{equation}
The orbital period $P$ and the orbital precession $\Delta\phi$ over one period are
\begin{equation}
    \label{period}
    P=\int_{0}^{2\pi}\frac{dt}{d\chi}d\chi,
\end{equation}
\begin{equation}
\begin{split}
    \label{phi} \Delta\phi=\int_{0}^{2\pi}\frac{d\phi}{d\chi}d\chi-2\pi.
\end{split}
\end{equation}

\begin{figure*}[htp]
    \centering
    \includegraphics[width=0.87\linewidth]{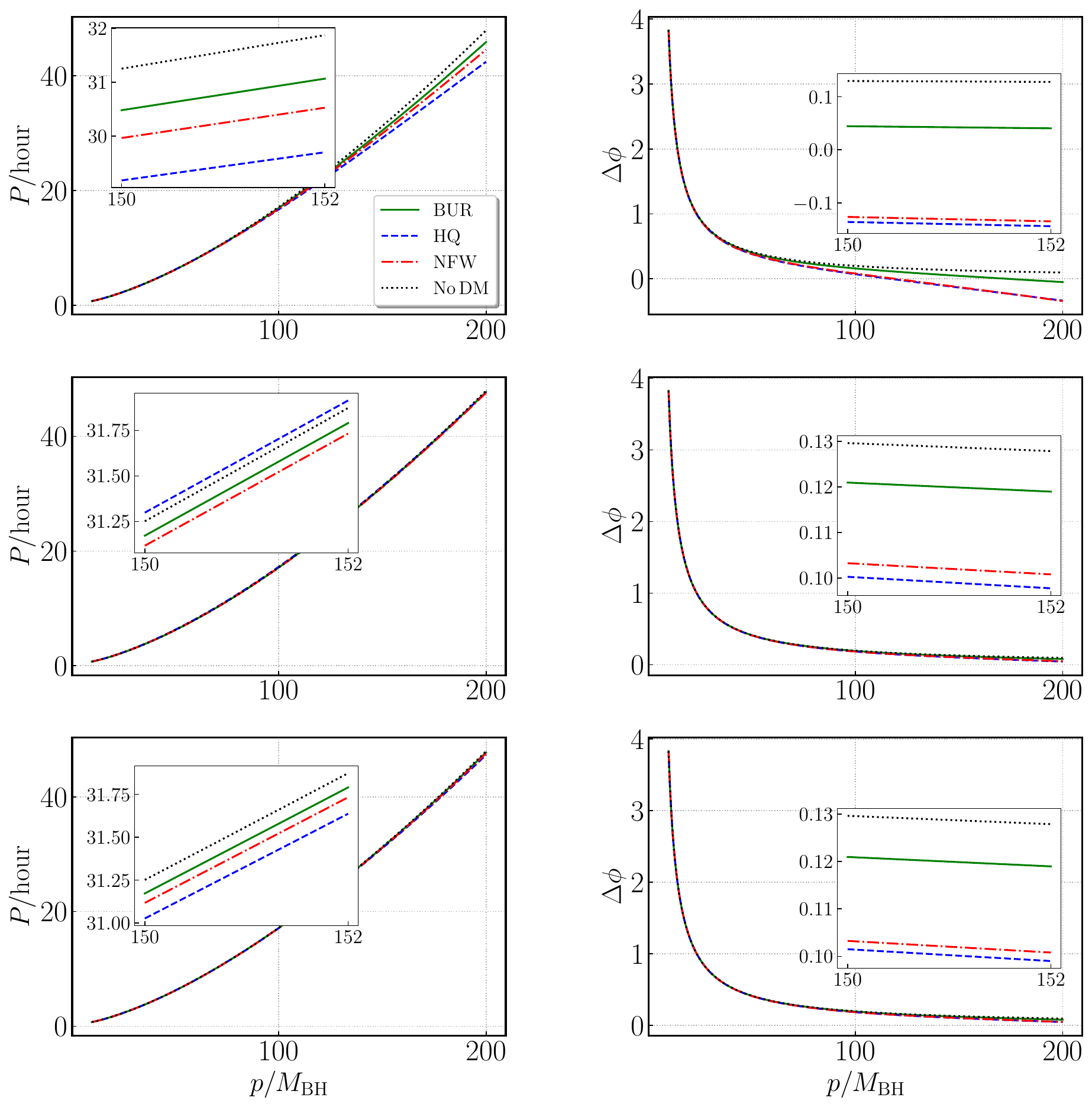}
    \caption{
    The results of the orbital periods and precessions for EMRIs in galaxies with different DM halos models. 
    The symbols ``BUR", ``HQ", ``NFW" and ``No DM" represent the results for the Burkert, Hernquist, NFW density distributions, as well as the case without DM halos, respectively.
    The mass of central SMBH $M_\text{BH}$ is $10^6M_\odot$,
    and the eccentricity $e=0.6$.
    The parameters $(a_0,M)$ are chosen as $(100M,100M_\text{BH})$, $(100M,1000M_\text{BH})$, and $(1000M,10M_\text{BH})$ in the top, middle and bottom panels, respectively.
    }
\label{fig:2}
\end{figure*}

By combining Eqs. \eqref{rpe}, \eqref{E-L1} and \eqref{orbital3}-\eqref{phi}, we numerically calculate the orbital periods and precessions of EMRIs and present them in Fig. \ref{fig:2}. 
From Fig. \ref{fig:2}, we observe that the periods increase with $p$, while the orbital precessions decrease with $p$. 
In the right panels of Fig. \ref{fig:2}, we also see that the orbital precessions of EMRIs within DM halos are smaller than those in vacuum cases. 
For the scenarios with $a_0=100M$ and $M=100M_\text{BH}$, the direction of precession changes from prograde to retrograde for all three DM halo types.
Comparing the top panels with the middle panels, 
we find that the differences in the orbital periods and precessions among the three DM halo-type EMRIs decrease as $M/a_0^2$ decreases. 
As $M/a_0$ is fixed and $M/a_0^2$ decreases, 
the period for the Hernquist model increases, 
becoming even larger than that without DM.
Additionally, from the middle and bottom panels, we see that orbital periods and precessions among the Burkert and NFW DM densities are nearly identical, as $M/a_0^2$ remains the same. 

\begin{figure*}[htp]
    \centering
    \includegraphics[width=0.87\linewidth]{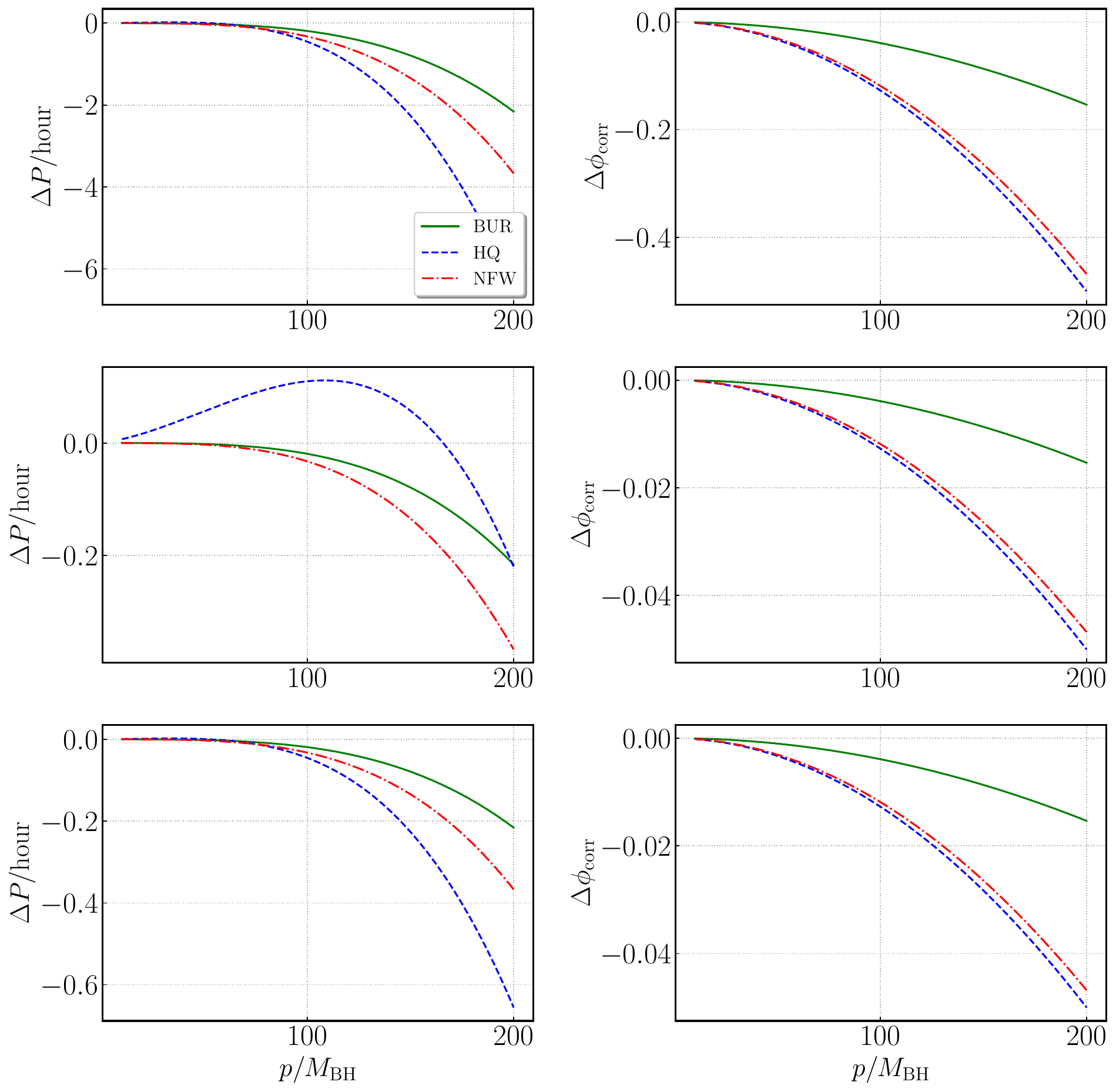}
    \caption{
    The corrections to the orbital periods and precessions due to DM halos. 
    The symbols ``BUR", ``HQ", and ``NFW" represent the results for the Burkert, Hernquist, and NFW density distributions.
    The mass $M_\text{BH}$ of central SMBH is $10^6M_\odot$,
    and the eccentricity $e=0.6$.
    The parameters $(a_0,M)$ are chosen as $(100M,100M_\text{BH})$, $(100M,1000M_\text{BH})$, and $(1000M,10M_\text{BH})$ in the top, middle, and bottom panels, respectively.
    }
\label{fig:2a}
\end{figure*}

To understand these numerical results, we derive approximate analytical formulas for the orbital period and precession in appendix \ref{APPENDIX1}, treating the compactness $M/a_0$ as a small quantity.
Using the approximate analytical formulas, 
we plot the corrections to the orbital period and precession due to DM halos in Fig. \ref{fig:2a}. 
For the Hernquist model, from Eq. \eqref{A-hq}, we see that the term $1-2M/a_0+4M^2/(3a_0^2)$ which depends on $M/a_0$ decreases the gravitational force exerted by the central SMBH, while the term proportional to $Mr/a_0^2$ due to DM halos enhances the gravitational force acting on the SCO,
leading to the positive corrections to $P$ which depend on the compactness $M/a_0$ and the negative corrections to $P$ which depend on $M/a_0^2$
in Eq. \eqref{HQP}.
For the Burkert and NFW models,
Eqs. \eqref{A-bur}, \eqref{BURP}, \eqref{A-nfw} and \eqref{NFWP} tell us that the presence of both Burkert-type and NFW-type DM halos enhances the gravitational force acting on the SCO and henceforth decreases the period $P$, 
with the decrease depending on $M/a_0^2$;
the larger the value of $M/a_0^2$, the bigger the decrease as shown in Figs. \ref{fig:2} and \ref{fig:2a}.
From Eqs. \eqref{HQPhi}, \eqref{BURPhi} and \eqref{NFWPhi},
we see that the presence of DM halos reduces the orbital precession or even reverses its direction, with the reduction depending on the density $M/a_0^2$ because the corrections to orbital precession due to DM halos are caused by the $Mr/a_0^2$ terms in Eqs. \eqref{A-hq}, \eqref{A-bur} and \eqref{A-nfw} only;
the larger the value of $M/a_0^2$, the bigger the correction as shown in Figs. \ref{fig:2} and \ref{fig:2a}.

EMRIs within DM halos radiate GWs and interact with the surrounding environment. 
The motion of the SCO is influenced by GW reaction, dynamic friction, and accretion from the halo medium. 
When the SCO is a small BH, it will accrete the surrounding DM as it moves through the halo, which can be described by the Bondi-Hoyle accretion model \cite{Bondi:1952ni, Bondi:1944jm,Macedo:2013qea}
\begin{equation}
\label{dmdt}
    \dot{\mu} = \frac{4 \pi \mu^2 \rho_\text{DM}}{(v^2+c_\text{s}^2)^{3/2}},
\end{equation}
where $c_\text{s}=\sqrt{\delta P_\text{t}/\delta \rho_\text{DM}}$ is the sound speed of the medium, $v$ is the velocity of the small BH, and the overdot denotes differentiation with respect to time.

The accretion does not exchange the orbital angular momentum and keeps the orbital shape unchanged, i.e. $\left({dL}/{dt}\right)_\text{acc}=0$
and $de/dt=0$ \cite{Hughes:2018qxz}, 
so the change in orbital energy caused by accretion can be written as \cite{Hughes:2018qxz}
\begin{equation}
\label{dEdtacc}
    \left(\frac{dE}{dt}\right)_\text{acc} =\frac{\dot{\mu}}{\mu} E+\mu\dot{\epsilon}= \frac{\dot{\mu}}{\mu} \left(E-{h}\frac{dp}{dh}\frac{dE}{dp} \right).
\end{equation}

The small BH interacts with the surrounding DM particles, resulting in a drag force known as dynamic friction, which is given by \cite{Chandrasekhar:1943ys,Cardoso:2020iji}
\begin{equation}
\label{fDF}
    \bm{f}_\text{DF}=-\frac{4\pi \mu^2 \rho_\text{DM} \ln\Lambda}{v^3}\bm{v},
\end{equation}
where the Coulomb logarithm $\ln\Lambda=3$ \cite{Eda:2014kra}.
The loss rates of orbital energy and angular momentum due to dynamical friction are given by
\begin{equation}
\label{dEdtdf}
     \left(\frac{dE}{dt}\right)_{\text{DF}}=\bm{f}_{\text{DF}}\cdot\bm{v},
\end{equation}
\begin{equation}
\label{dLdtdf}
    \left(\frac{dL}{dt}\right)_{\text{DF}}=\bm{r}\times \bm{f}_{\text{DF}}.
\end{equation}

The energy and angular momentum fluxes due to the GWs radiation of the quadrupole form are \cite{Peters:1963ux}
\begin{equation}
\label{dEdtGW1}
    \left(\frac{dE}{dt}\right)_{\text{GW}}=-\frac{1}{5}\dddot{\mathcal{I}}^{jk}\dddot{\mathcal{I}}^{jk},
\end{equation}
\begin{equation}
\label{dLdtGW1}
    \left(\frac{dL_i}{dt}\right)_{\text{GW}}=-\frac{2}{5}\epsilon_{ijk}\ddot{\mathcal{I}}^{jl}\dddot{\mathcal{I}}^{kl},
\end{equation}
where the symmetric tracefree form of the mass quadrupole is
\begin{equation}
\label{quadrupole-st}
    \mathcal{I}^{jk}=\mu(t) \left(x^jx^k-\frac{1}{3}r^2\delta^{jk}\right),
\end{equation}
$x^j=(r\cos\phi,\,r\sin\phi,\,0)$. 

The changes in orbital energy and angular momentum resulting from the combined effects of GWs radiation, accretion, and dynamical friction can be expressed as 
\begin{equation}
\label{dEdtorb}
    \left(\frac{dE}{dt}\right)_{\text{orb}}=\left<\frac{dE}{dt}\right>_{\text{GW}}+\left<\frac{dE}{dt}\right>_{\text{acc}}+\left<\frac{dE}{dt}\right>_{\text{DF}},
\end{equation}
\begin{equation}
\label{dLdtorb}
    \left(\frac{dL}{dt}\right)_{\text{orb}}=\left<\frac{dL}{dt}\right>_{\text{GW}}+\left<\frac{dL}{dt}\right>_{\text{acc}}+\left<\frac{dL}{dt}\right>_{\text{DF}},
\end{equation}
where the angle brackets denote averaging over several gravitational wavelengths, and
\begin{equation}
\begin{split}
\label{averesult1}
    \left<\frac{dE}{dt}\right>=\frac{1}{P}\int_{0}^{P}\frac{dE}{dt}dt=\frac{1}{P}\int_{0}^{2\pi}\frac{dE}{dt}\frac{dt}{d\phi}d\phi,
\end{split}
\end{equation}
\begin{equation}
\begin{split}
\label{averesult2}
    \left<\frac{dL}{dt}\right>=\frac{1}{P}\int_{0}^{P}\frac{dL}{dt}dt=\frac{1}{P}\int_{0}^{2\pi}\frac{dL}{dt}\frac{dt}{d\phi}d\phi.
\end{split}
\end{equation}

\begin{figure*}[htp]
    \centering
    \includegraphics[width=0.95\linewidth]{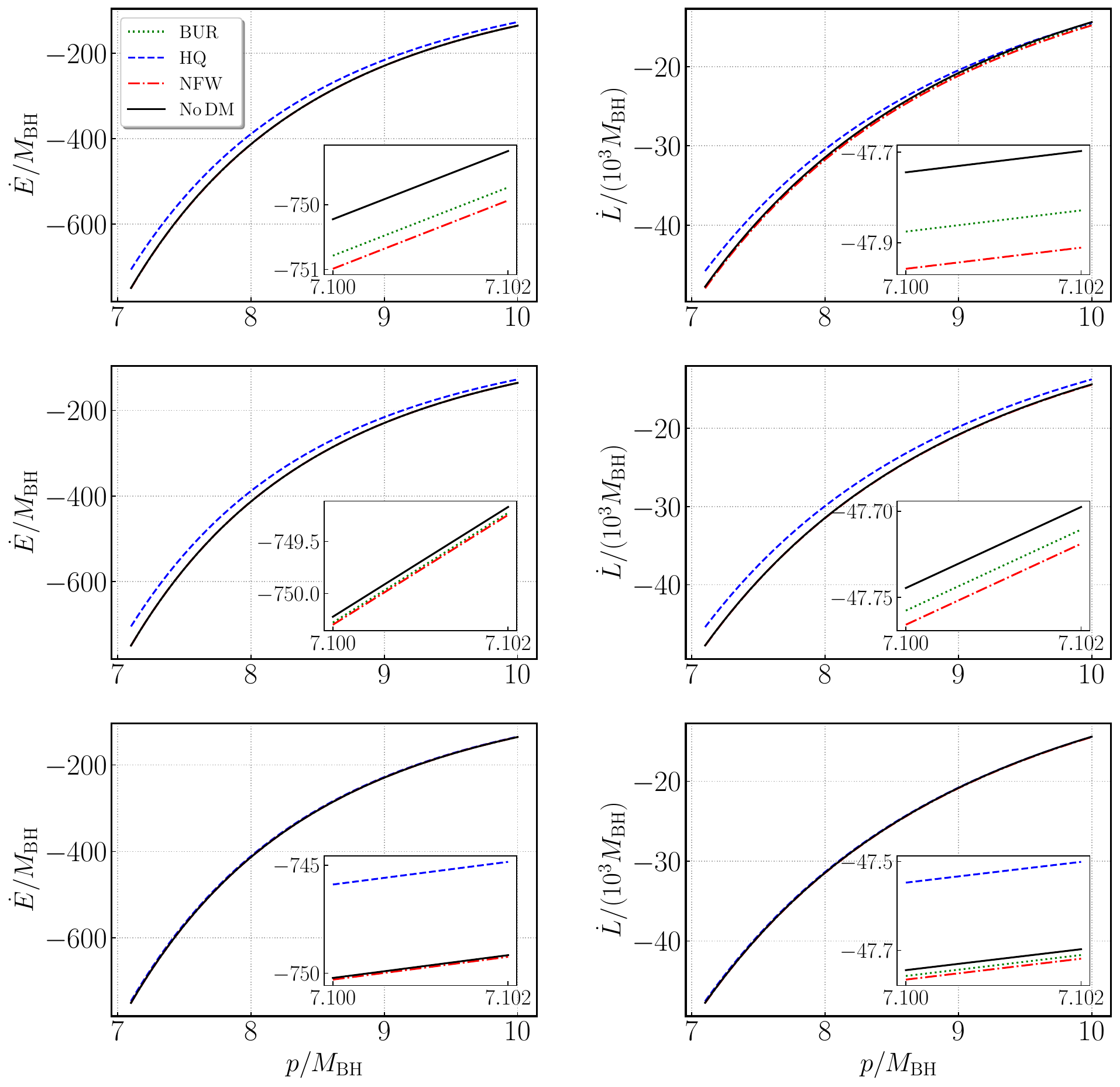}
    \caption{
    The energy and angular momentum fluxes for EMRIs in galaxies with  three specific types of DM halos and without DM halos.
    The mass of central SMBH is $M_\text{BH}=10^6M_\odot$, the mass of the small BH is $\mu=10 M_\odot$, and the eccentricity is $e=0.6$.
    The parameters $(a_0,M)$ are chosen as $(100M,100M_\text{BH})$, $(100M,1000M_\text{BH})$, and $(1000M,10M_\text{BH})$ in the top, middle, and bottom panels, respectively. 
    }
\label{fig:3}
\end{figure*}

By combining Eqs. \eqref{dEdtacc}, \eqref{dEdtdf}-\eqref{dLdtGW1}, \eqref{dEdtorb}, and \eqref{dLdtorb}, we numerically calculate the energy and angular momentum fluxes resulting from the combined effects of GW reactions, accretion, and dynamic friction, and the results are shown in Fig. \ref{fig:3}. 
From Fig. \ref{fig:3}, we observe that the fluxes generated by the three distinct types of DM halos differ from those in scenarios without DM. 
When $p<10 M_\text{BH}$, the energy loss rate for Hernquist-type DM halos is lower than in cases without DM halos, whereas the losses increase for NFW-type and Burkert-type DM halos.

\begin{figure*}[htp]
    \centering  
    \includegraphics[width=0.9\linewidth]{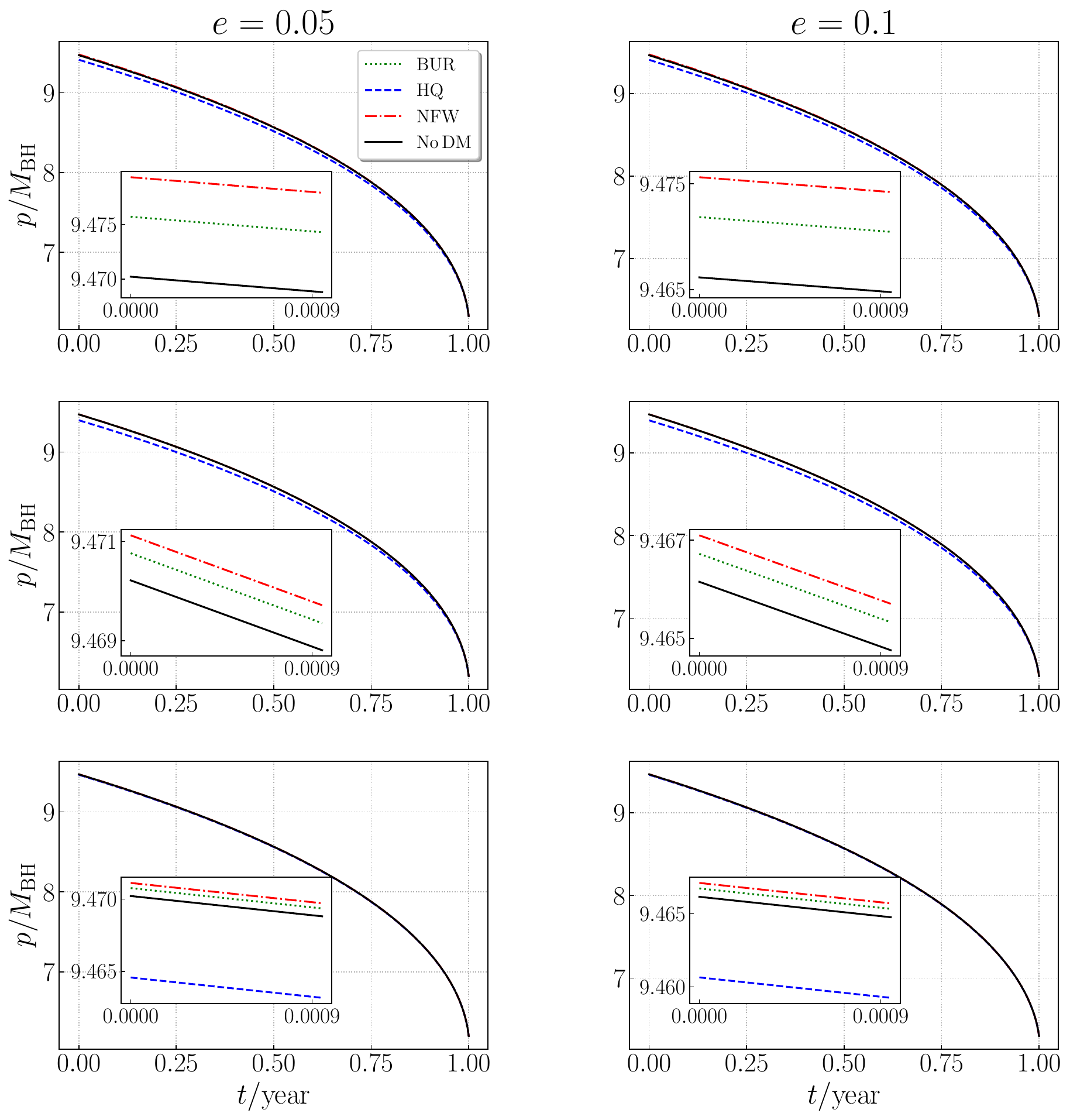}
    \caption{
    The evolution of $p(t)$ with three types of the DM halos and without DM halos for one year before the last stable orbit.
    The mass of central SMBH is $M_\text{BH}=10^6M_\odot$ and the mass of the small BH is $\mu=10 M_\odot$.
    The eccentricities at the last stable orbit are $e=0.05$ and $e=0.1$.
    The parameters $(a_0,M)$ are chosen as $(100M,100M_\text{BH})$, $(100M,1000M_\text{BH})$, and $(1000M,10M_\text{BH})$ in the top, middle, and bottom panels, respectively. 
    }
\label{fig:4}
\end{figure*}

\begin{figure*}[htp]
    \centering
    \includegraphics[width=0.95\linewidth]{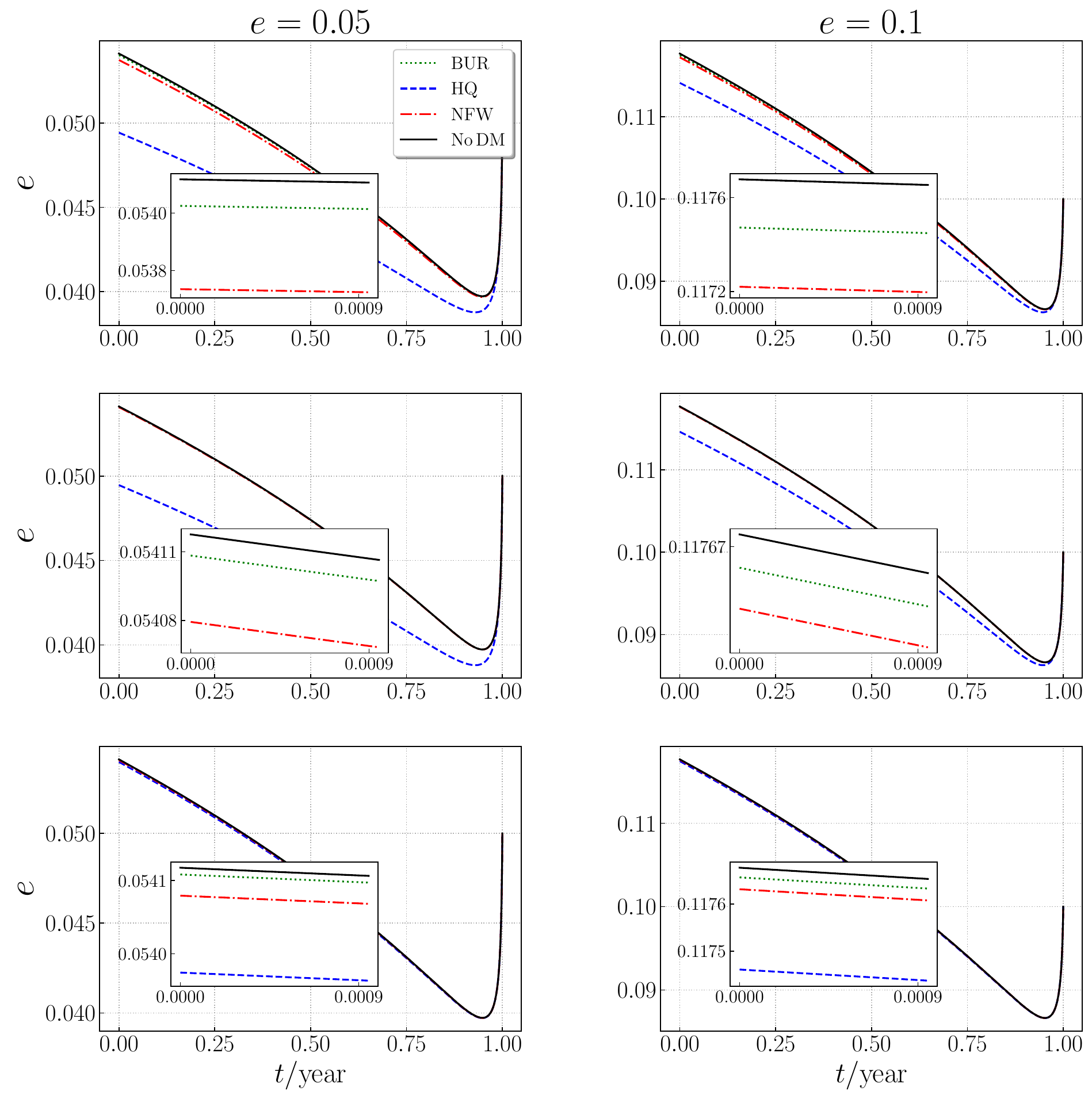}
    \caption{
    The evolution of eccentricity $e(t)$ with three types of the DM halos and without DM halos for one year before the last stable orbit.
    The mass of central SMBH is $M_\text{BH}=10^6M_\odot$ and the mass of the SCO is $\mu=10 M_\odot$.
    The eccentricities at the last stable orbit are $e=0.05$ and $e=0.1$.
    The parameters $(a_0,M)$ are chosen as $(100M,100M_\text{BH})$, $(100M,1000M_\text{BH})$, and $(1000M,10M_\text{BH})$ in the top, middle, and bottom panels, respectively. 
    }
\label{fig:5}
\end{figure*}

Combining Eqs. \eqref{E-L1}, \eqref{dEdtacc}, \eqref{dEdtdf}-\eqref{dLdtGW1}, \eqref{dEdtorb}, and \eqref{dLdtorb}, 
we numerically calculate the evolution of $p(t)$ and $e(t)$, with the results shown in Figs. \ref{fig:4} and \ref{fig:5}. 
In Fig. \ref{fig:4}, 
we observe that the evolution of EMRIs within Hernquist-type DM halos significantly affects the semi-latus rectum and eccentricity, 
while the Burkert density distribution has the least influence. 
This is because that the density of the Hernquist-type DM halos is the largest, while the Burkert-type DM halos is the smallest.
EMRIs embedded in Hernquist-type DM halos evolve more slowly compared to those without DM halos, 
whereas EMRIs in Burkert-type and NFW-type halos evolve more rapidly. 
This is because a greater amount of energy loss leads to quicker evolution, 
and the energy loss rate for Hernquist-type DM halos is lower than in cases without DM halos, whereas the losses increase for other two models.
From Fig. \ref{fig:5}, we see that the presence of DM halos reduces the rate of eccentricity decrease for EMRIs across all three types of DM halos.

By comparing the top and middle panels in Figs. \ref{fig:4} and \ref{fig:5}, 
with $M/a_0$ held constant, we observe that larger values of $M/a_0^2$ lead to greater differences in the evolution of the semi-latus rectum and eccentricity across the three DM halo models. 
Similarly, comparing the middle and bottom panels, with $M/a_0^2$ fixed, shows that larger values of $M/a_0$ result in more pronounced differences in these orbital parameters. 
From Figs. \ref{fig:4} and \ref{fig:5}, we conclude that higher compactness and density distribution of DM halos lead to larger variations in the evolution of both the semi-latus rectum and eccentricity.

\section{Gravitational waveforms}
\label{waveform}

\begin{figure*}[htp]
    \centering
   \includegraphics[width=0.95\linewidth]{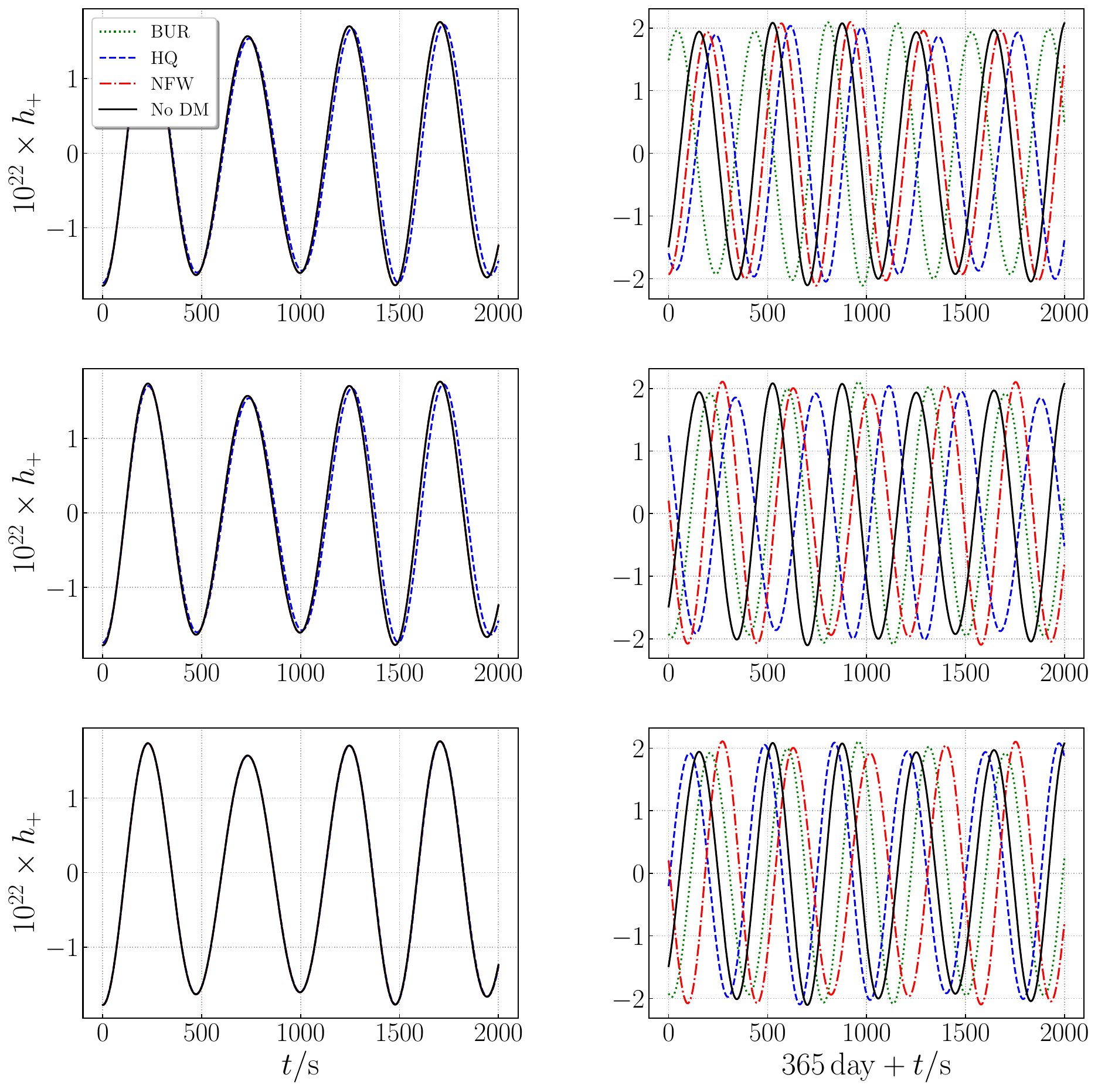}
    \caption{
    The time-domain plus mode waveforms for EMRIs within the three types of DM halos.
    The left panels show the waveforms at the beginning, and the right panels show the waveforms after one-year of evolution.
    We choose $M_\text{BH}=10^6M_\odot$, and $\mu=10 M_\odot$.
    The parameters $(a_0,M)$ are chosen as $(100M,100M_\text{BH})$, $(100M,1000M_\text{BH})$, and $(1000M,10M_\text{BH})$ in the top panels, middle panels, and bottom panels, respectively. 
    We take the initial semi-latus rectum $p_0=10M_\text{BH}$, the initial eccentricity $e_0=0.05$,
    the inclination angle $\iota=\pi/6$, the luminosity distance  $d_\text{L}=1\,\text{Gpc}$, and the initial longitude of pericenter $\omega_0=0$.
    }
\label{fig:6}
\end{figure*}

The presence of DM halos will influence the orbital motion of EMRIs, which will be reflected in the GWs emitted by them. 
The quadrupole formula for GWs is given by
\begin{equation}
\label{h-jk}
h^{jk}=\frac{2}{d_\text{L}} {\ddot{I}}^{jk},
\end{equation}
where $d_\text{L}$ is the luminosity distance from the detector to source, and $I^{jk}=\mu x^j x^k$ is the quadrupole moment.
When accounting for the effects of accretion, Eq. \eqref{h-jk} can be rewritten as
\begin{equation}
\label{hjkt}
\begin{split}
h^{jk}&=\frac{2}{d_\text{L}} \bigg\{\dot{\mu}(t)\left[4 r\dot{r}n^jn^k+2\dot{\phi}r^2(\lambda^jn^k+\lambda^kn^j)\right]\\
&+2\mu(t)\left[(\dot{r}^2+(\ddot{r}-r\dot{\phi}^2) r)n^jn^k+r\dot{r}\dot{\phi}(n^j\lambda^k+n^k\lambda^j)+r^2\dot{\phi}^2\lambda^j\lambda^j\right]+\ddot{\mu}(t)r^2n^jn^k\bigg\},
\end{split}
\end{equation}
where $n^j=(\cos\phi,\, \sin\phi,\, 0)$ and $\lambda^j=(-\sin\phi,\, \cos\phi,\, 0)$ are a couple of basis vectors in the orbital plane.
In the absence of DM halos ($M=0$), Eq. \eqref{hjkt} reduces to the standard quadrupole formula \cite{Peters:1963ux}.
The plus and cross polarization modes of GWs are
\begin{equation}
    h_+=\frac{1}{2}(e_\text{X}^je_\text{X}^k-e_\text{Y}^je_\text{Y}^k)h_{jk},
\end{equation}
\begin{equation}
    h_\times=\frac{1}{2}(e_\text{X}^je_\text{Y}^k+e_\text{Y}^je_\text{X}^k)h_\text{jk},
\end{equation}
where $e_\text{X}$ and $e_\text{Y}$ are orthonormal vectors in the plane perpendicular to the direction from the detector to the GW source.
We present the plus-mode GWs in Fig. \ref{fig:6}. 
Initially, the waveforms of EMRIs within the three types of DM halos are difficult to distinguish.
However, after one year of evolution, the waveforms of EMRIs in different types of DM halo become clearly distinguishable.

\begin{figure*}[htp]
    \centering{$
    \begin{array}{cc}
    \includegraphics[width=0.45\textwidth]{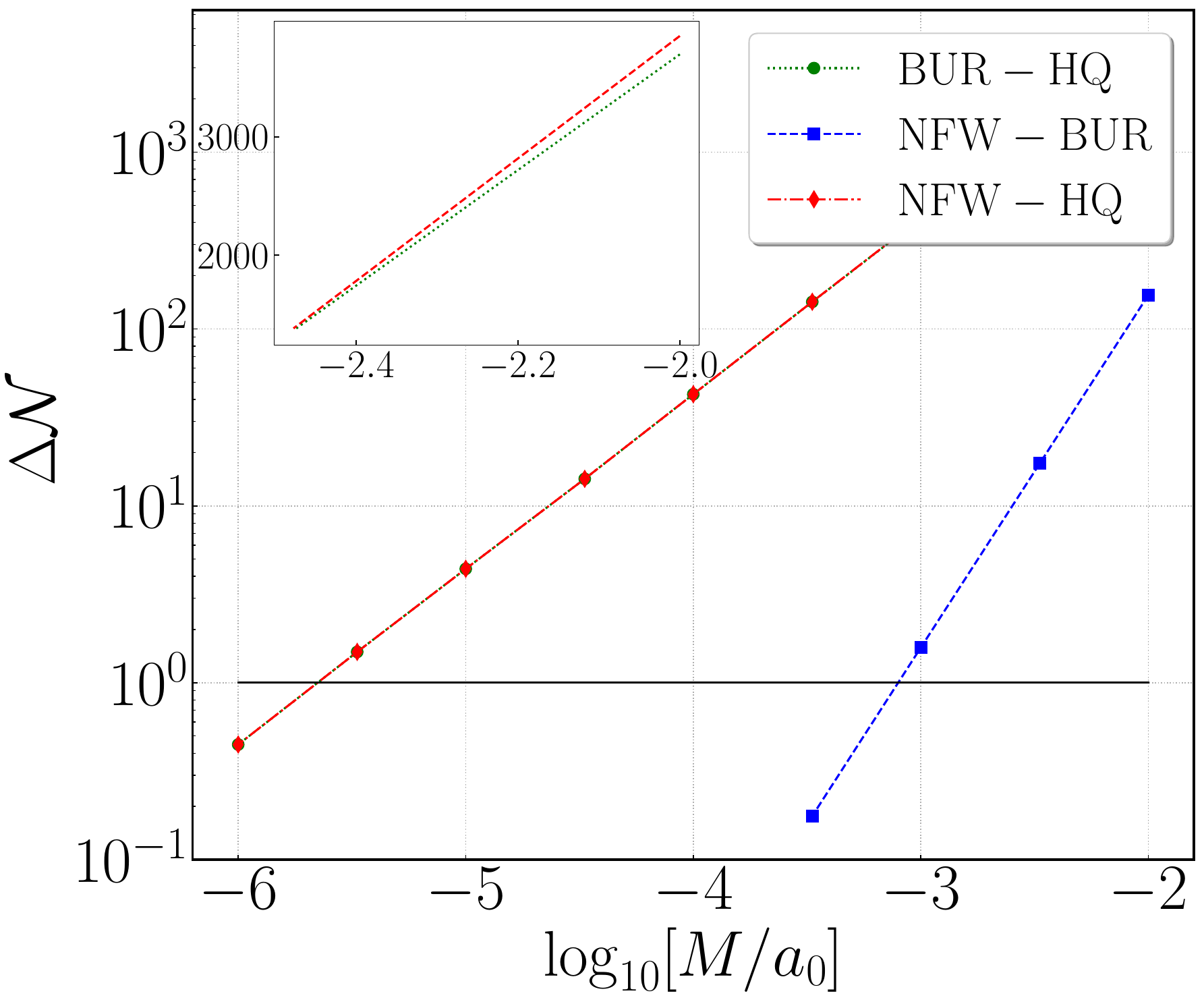}\quad \quad &
    \includegraphics[width=0.45\textwidth]{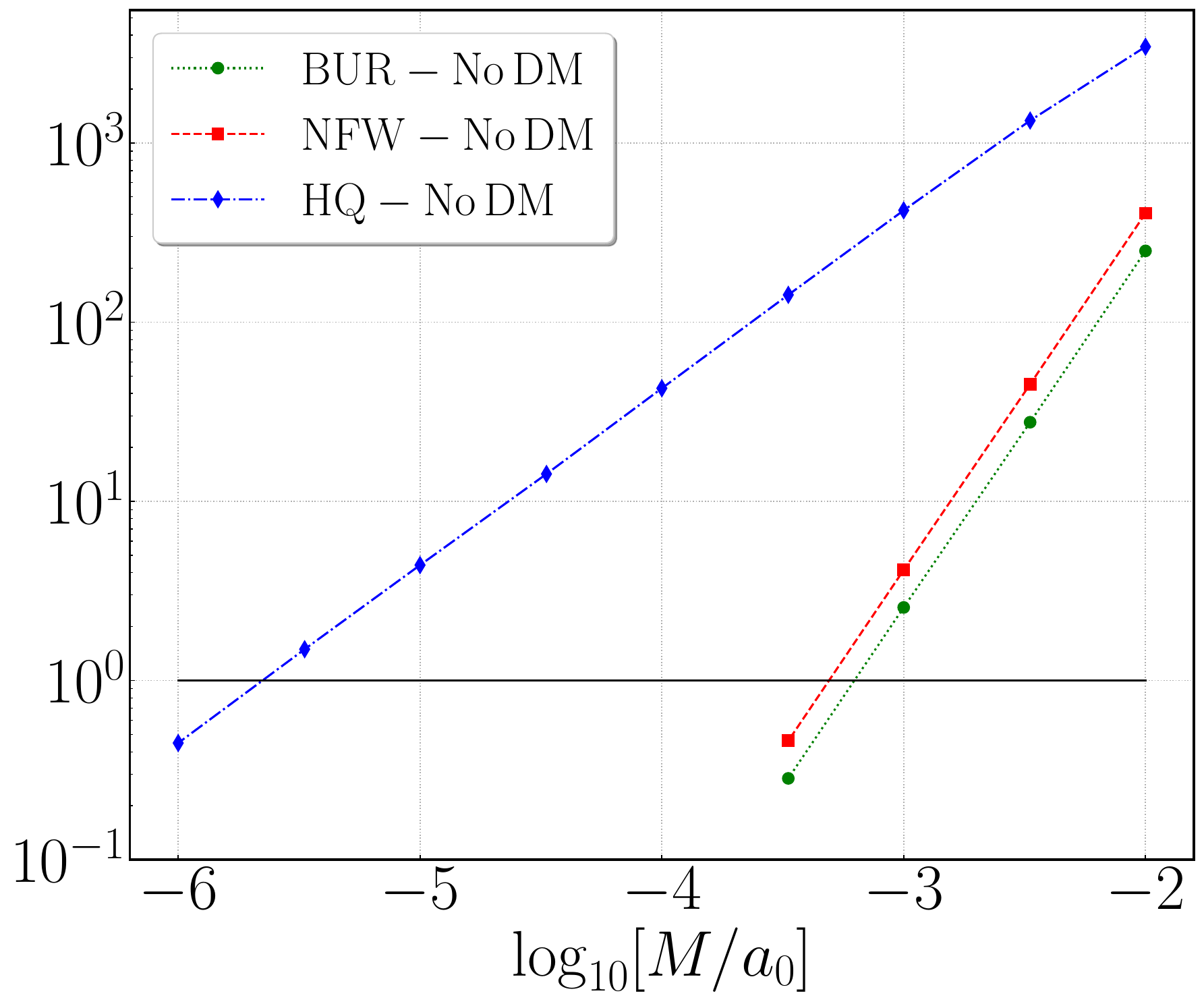}
    \end{array}$}
    \caption{
    The difference between the orbital cycles $\Delta\mathcal{N}$ for EMRIs accumulated over one-year evolution.
    The left panel shows the differences in orbital cycles of EMRIs across different types of DM halos.
    The right panel shows the differences in orbital cycles of EMRIs within DM halos compared to those without DM.
    The initial semi-latus rectum $p_0$ is chosen at the position where the orbital period $\text{T}=2\pi\sqrt{2(5 R_\text{s})^3/R_\text{s}}$.
    We choose the initial eccentricity $e_0=0.05$, 
     $M_\text{BH}=10^6M_\odot$, $\mu=10 M_\odot$, and $M=100M_\text{BH}$. 
    The black line marks $\Delta\mathcal{N}=1$ rad. 
    }
\label{fig:7}
\end{figure*}

To quantitatively analyze the effects of different halo models on the evolution of EMRIs, we calculate the number of orbital cycles for EMRIs within DM halos. 
The number of orbital cycles is given by \cite{Maselli:2020zgv, Barsanti:2022ana} 
\begin{equation}
\label{circle}
    \mathcal{N}=\int_{t_1}^{t_2}\dot{\phi}(t)\,dt.
\end{equation}
The difference in the number of orbital cycles between two EMRIs in different types of DM halos can be expressed as $\Delta \mathcal{N}=\mathcal{N}_1-\mathcal{N}_2$. 
Combining Eqs. \eqref{orbital3}, \eqref{orbital2}, and \eqref{circle}, the orbital cycles can be computed numerically.
In Fig. \ref{fig:7}, we show the difference in the number of orbital cycles for EMRIs after one year of evolution. 
Following \cite{Maselli:2020zgv}, we adopt $\Delta\mathcal{N} \sim 1$ rad as the threshold for detectable dephasing. 

In the left panel of Fig. \ref{fig:7}, we observe that Burkert-type and NFW-type DM halos can be distinguished when $M/a_0>10^{-3}$. 
Similarly, NFW-type and Hernquist-type halos, as well as Burkert-type and Hernquist-type halos, become distinguishable when $M/a_0 >10^{-5}$. 
From the right panel, we see that Hernquist-type halos are detectable when $M/a_0>10^{-5}$, 
while Burkert-type and NFW-type halos are identifiable when $M/a_0>10^{-3}$.

The threshold for orbital cycles cannot be regarded as a sufficient condition for detectability. 
To more accurately distinguish between different waveforms, we calculate the mismatch of GWs signals.
The GWs measured by the detector are given by
\begin{equation}
\label{ht}
    h(t)=h_+(t)F^++h_\times(t) F^\times,
\end{equation}
where $F^+$ and $F^\times$ are the detector pattern functions \cite{Barack:2003fp}.
The inner product between two waveforms $h_1$ and $h_2$ are
\begin{equation}
    \langle h_1\vert h_2 \rangle=2\int_{f_\text{min}}^{f_\text{max}}\frac{\widetilde{h}_1(f)\widetilde{h}_2^*(f)+\widetilde{h}_2(f)\widetilde{h}_1^*(f)}{S_n(f)}df,
\end{equation}
where $f_\text{min}$ and $f_\text{max}$ are
\begin{equation}
\begin{split}
    f_\text{min}&=\text{Min}(f_\text{end},f_\text{up}),\\
    f_\text{max}&=\text {Max}(f_\text{ini},f_\text{low}),
\end{split}
\end{equation}
$f_\text{ini}$ and $f_\text{end}$ are the initial and final frequencies for the orbital evolution, and the lower and upper cutoff frequencies for LISA are chosen as $f_\text{low}=10^{-4}\text{Hz}$ and $f_\text{up}=1\text{Hz}$, respectively \cite{Yagi:2009zm}.
$\widetilde{h}(f)$ is the Fourier transformation of the time-domain signal $h(t)$, and $\widetilde{h}^*(f)$ is its complex conjugate. 
$S_n(f)$ is the noise spectral density for GWs detectors. 
The one-side noise power spectral density of LISA is \cite{Robson:2018ifk}
\begin{equation}
    \label{psd-lisa}
    S_\text{n}(f) =\frac{S_x}{L^2}+\frac{2S_a \left[1+\cos^2(2\pi f L/c)\right]}{(2\pi f)^4 L^2}\times\left[1+\left(\frac{4\times 10^{-4}\text{Hz}}{f}\right) \right].
\end{equation}
For LISA, the length of the detector arm is $L=2.5\times 10^9\text{m}$, the displacement noise is $\sqrt{S_\text{x}}=1.5\times 10^{-11}\text{m}\,\text{Hz}^{-1/2}$ and the acceleration noise is $\sqrt{S_a}=3\times 10^{-15} \text{m}\,\text{s}^{-2}\,\text{Hz}^{-1/2}$.

The signal-to-noise ratio (SNR) is $\text{SNR} = \left< h\vert h \right>$.
The faithfulness can be written as
\begin{equation}
\label{faithfulness}
    \mathcal{F}[h_1,h_2]=\text{Max}_{(t_0,\phi_0)}\frac{\left< h_1|h_2 \right>}{\sqrt{\left< h_1\vert h_1 \right> \left< h_2\vert h_2 \right>}},
\end{equation}
where the $(t_0,\phi_0)$ are the time and phase offsets.
The mismatch between two signals is
\begin{equation}
\label{Mismatch}
\text{Mismatch}[h_1, h_2] = 1-\mathcal{F}[h_1, h_2].
\end{equation}
The two waveforms can be distinguished only when $\text{Mismatch}[h_1, h_2]>\text{d}/(2\text{SNR}^2)$ is satisfied, 
where $d$ represents the number of parameters for the GW sources, 
and $d=13$ for EMRIs within DM halos \cite{Flanagan:1997kp, Lindblom:2008cm}.

\begin{figure*}[htp]
    \centering{$
    \begin{array}{cc}
    \includegraphics[width=0.45\textwidth]{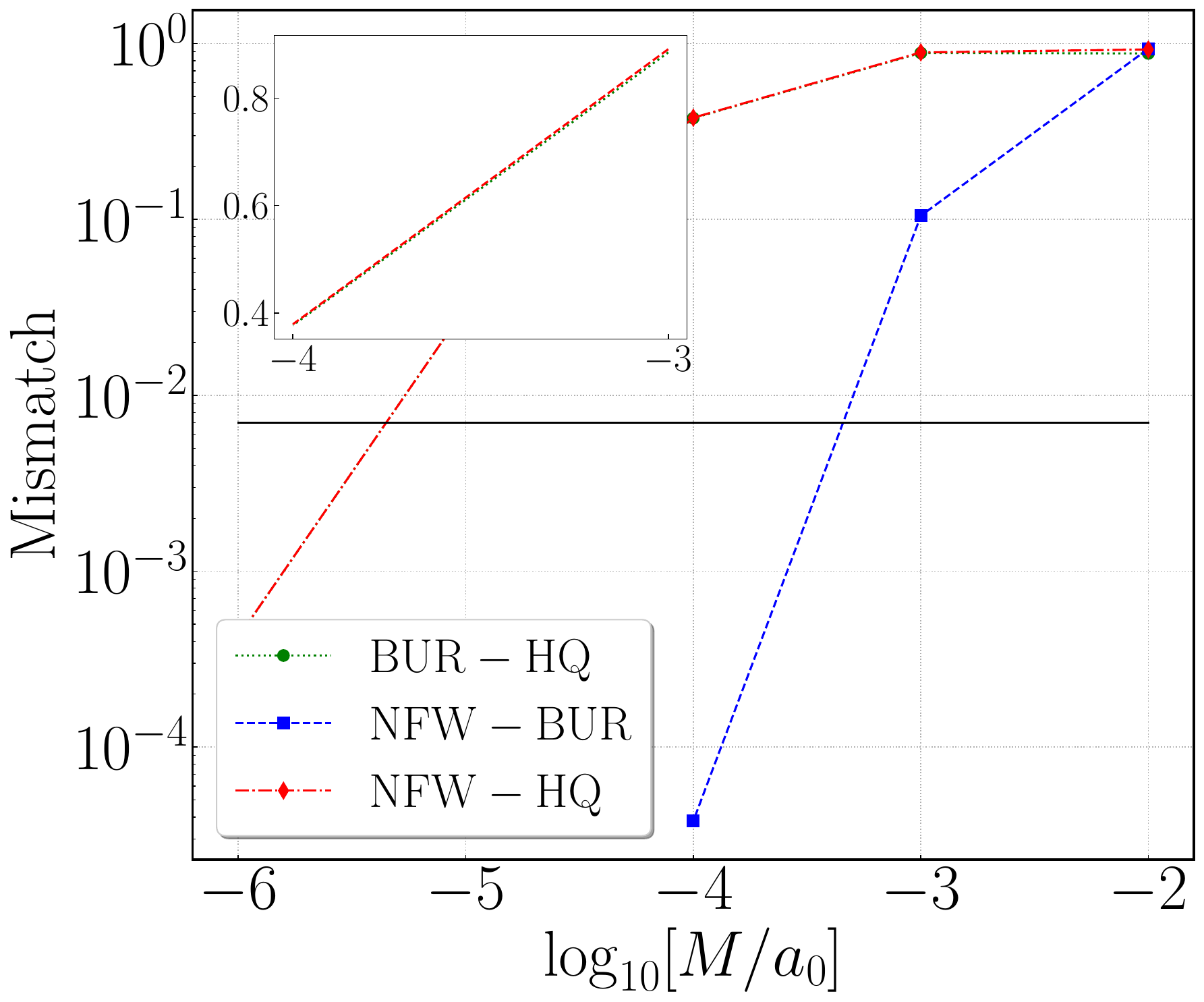}\quad \quad &
    \includegraphics[width=0.45\textwidth]{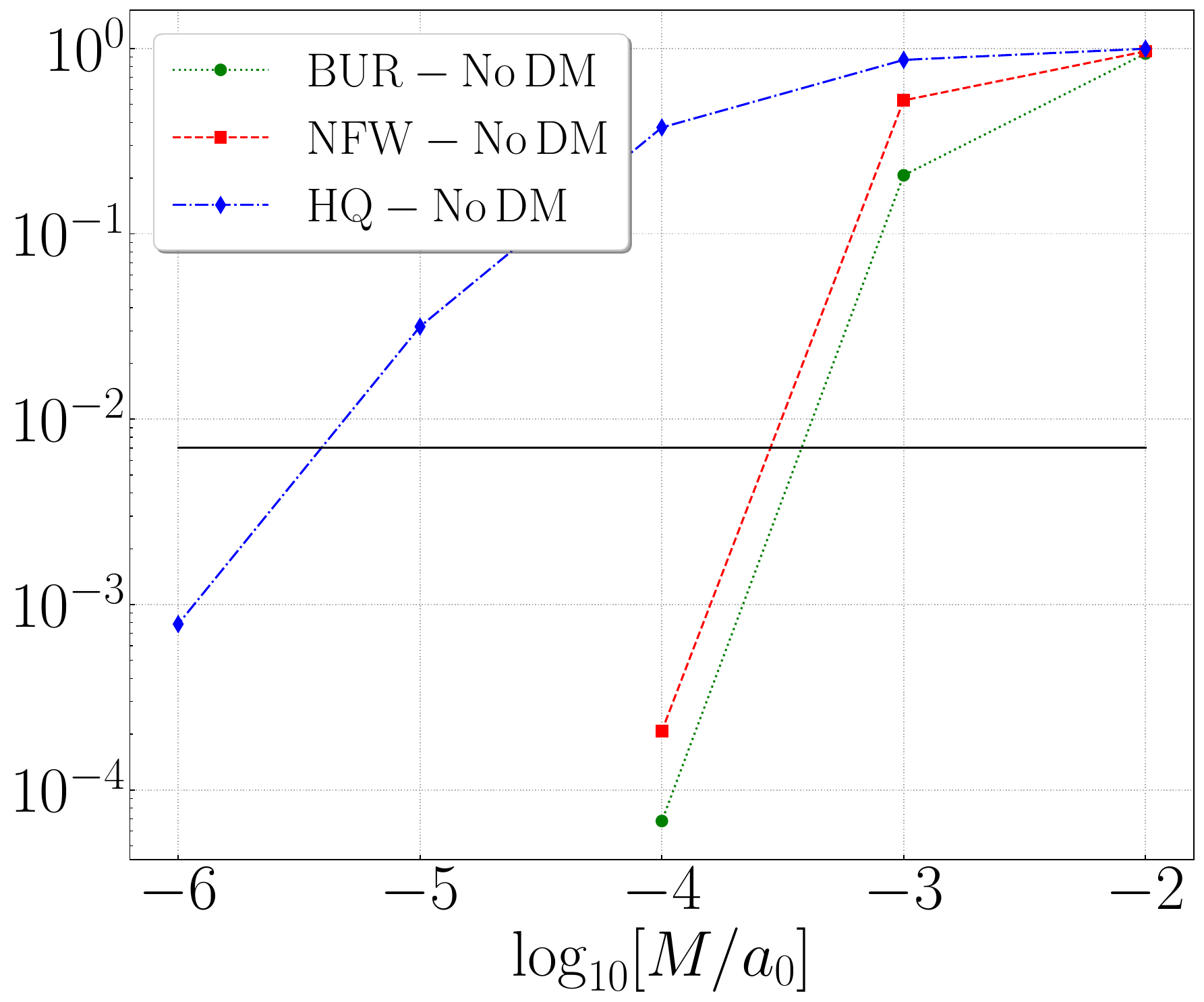}
    \end{array}$}
    \caption{
    The mismatch between GWs for one-year evolution before merger.
    The left panel shows the mismatch between GWs emitted from EMRIs within different types of DM halos.
    The right panel shows the mismatch between GWs with and without DM halos.
    The black line is the threshold $d/\text{SNR}^2\simeq 0.007$.
    The initial eccentricity $e_0=0.05$, $M_\text{BH}=10^6M_\odot$, $\mu=10 M_\odot$, $M=100M_\text{BH}$,
    the luminosity distance $d_\text{L} = 1\text{Gpc}$, the inclination angle $\iota = \pi/6$ and the longitude of pericenter $\omega=0$.
    The SNRs of the GWs detected with LISA are about $32$.
    }
\label{fig:8}
\end{figure*}

Combining Eqs. \eqref{dEdtorb}-\eqref{averesult2}, \eqref{hjkt}, \eqref{ht}, \eqref{faithfulness}, and \eqref{Mismatch}, 
we numerically calculate the mismatch between GWs from EMRIs within the three types of DM halo, 
as well as the mismatch between GWs with and without DM halos. 
The results are presented in Fig. \ref{fig:8}.
In the left panel of Fig. \ref{fig:8}, we observe that the differences in GW emissions from EMRIs within Hernquist-type and NFW-type DM halos, 
as well as between Hernquist-type and Burkert-type DM halos, become distinguishable when $M/a_0>10^{-5}$. 
The distinction between GWs from EMRIs in Burkert-type and NFW-type DM halos becomes apparent when $M/a_0>10^{-3}$. 
In the right panel of Fig. \ref{fig:8}, we find that Hernquist-type DM halos are detectable when $M/a_0>10^{-5}$, 
while Burkert-type and NFW-type DM halos can be detected when the compactness satisfies $M/a_0>10^{-3}$. 
The Burkert-type DM halo is the most challenging to detect, whereas the Hernquist-type is the easiest. 
This is due to the fact that the density of the Hernquist-type DM halo is the highest, 
while the Burkert-type DM halo has the lowest density for the same values of $a_0$ and $M$. 
Lower-density DM halos are more difficult to detect.



\section{CONCLUSION AND DISCUSSION}
\label{conclusion}

Using the analytic, static, and spherically symmetric metric for a Schwarzschild BH immersed in DM halo, 
we investigated the feasibility for differentiating various DM halo models through EMRIs.
We calculated the orbital periods and precessions for eccentric EMRIs within the three DM halo types.
For the Hernquist model, we find that the gravitational force exerted by the central MBH is decreased by DM halos, while DM halos put additional gravitational force on the SCO.
The presence of both Burkert-type and NFW-type DM halos enhances the gravitational force acting on the SCO, 
resulting in a decrease in the period $P$, 
with the decrease depending on $M/a_0^2$;
the larger the value of $M/a_0^2$, the bigger the decrease.
Additionally, we find that the reduction in orbital precession due to DM halos is influenced by $M/a_0^2$;
the larger the value of $M/a_0^2$, the bigger the reduction.
The results indicate that DM halos reduce the precession rate of EMRIs, 
and the orbital precession may become retrograde if the local DM halo density is sufficiently high. 
Larger values of $M/a_0$ and $M/a_0^2$ lead to greater differences in the effects on orbital periods and precessions among the three DM halo models.

Considering accretion, dynamical friction, and GW reactions, we derived the loss rates of the orbital energy and angular momentum for the three types of DM halos. 
When $p<10 M_\text{BH}$, compared to the scenario without a DM halo, the orbital energy loss rate of EMRIs within Hernquist-type DM halos decreases, 
while it increases for those within Burkert-type and NFW-type halos. 
We numerically calculated the evolution of the semi-latus rectum $p(t)$ and the eccentricity $e(t)$. 
Our results show that EMRIs within Hernquist-type DM halos evolve more slowly compared to cases without DM halos, 
whereas the evolution of EMRIs within Burkert-type and NFW-type halos occurs more rapidly. 
Additionally, the presence of DM halos slows the rate of eccentricity decrease. 
Higher compactness and density of the DM halos result in more pronounced differences in the evolution of both the semi-latus rectum and eccentricity.

We calculate the differences in orbital cycles for EMRIs with and without DM halos, 
as well as the differences among EMRIs within the three types of DM halos, accumulated over one year of evolution. 
When comparing the number of orbital cycles between EMRIs with and without DM halos, 
we find that Hernquist-type DM halos are detectable when $M/a_0>10^{-5}$, 
while Burkert-type and NFW-type DM halos are detectable when $M/a_0>10^{-3}$.
When comparing the number of orbital cycles among the three types of DM halos, distinctions between NFW-type and Burkert-type halos can be made when $M/a_0>10^{-3}$. 
Additionally, for $M/a_0>10^{-5}$, we can differentiate between NFW-type and Hernquist-type halos, as well as between Burkert-type and Hernquist-type halos.
By calculating the GW mismatch between EMRIs with and without DM halos, 
we find that Hernquist-type halos are the easiest to detect, becoming detectable when $M/a_0>10^{-5}$. 
In contrast, Burkert-type halos are the most challenging to detect, though both Burkert-type and NFW-type halos become detectable when $M/a_0>10^{-3}$. 
When comparing the GW mismatch among the three types of DM halos, distinguishing between NFW-type and Hernquist-type halos is the easiest, achievable when $M/a_0>10^{-5}$. 
Distinguishing between NFW-type and Burkert-type halos is more difficult but can be accomplished when $M/a_0>10^{-3}$. 
We can also distinguish between Burkert-type and Hernquist-type halos when $M/a_0>10^{-5}$.

In this paper, both dynamical friction and accretion are modeled using Newtonian mechanics. 
However, the relativistic effects of dynamical friction and accretion \cite{Speeney:2022ryg,Traykova:2023qyv,Petrich:1988zz,Mach:2021zqe} on EMRIs require further study.
To obtain a self-consistent result of environmental effects, it is essential to account for halo feedback from both accretion and dynamical friction \cite{Kavanagh:2020cfn, Becker:2021ivq}. 
These aspects will be explored in future research.

\acknowledgments
The computing work in this paper is supported by the Public Service Platform of High Performance Computing by Network and Computing Center of HUST.
This research is supported in part by the National Key Research and Development Program of China under Grant No. 2020YFC2201504.

\appendix
\section{The approximate analytic formulas}
\label{APPENDIX1}
To understand the results obtained, we present some approximate analytic formulas in this Appendix.
In astrophysical scenarios, the typical compactness of DM halos is $M/a_0\lesssim 10^{-4}$,
so we expand $A(r)$ and $m(r)$ about $M/a_0=0$ to the second order and use the ansatz to solve Einstein equation; the results are
\begin{equation}
\label{A-hq}
    A_\text{HQ}(r)\approx \left(1-\frac{2M_{BH}}{r}\right)\left(1-\frac{2M}{a_0}+\frac{4M^2}{3a_0^2}+\frac{2Mr}{a_0^2}\right),
\end{equation}

\begin{equation}
\label{m-hq}
    m_\text{HQ}(r)\approx M_\text{BH}+\frac{M r^2}{a_0^2}\left(1-\frac{2M_\text{BH}}{r}\right)^2,
\end{equation}

\begin{equation}
\label{rho-hq}
    \rho_\text{HQ}(r)\approx \frac{M\,(r-2M_\text{BH})}{2\pi r^2 a_0^2},
\end{equation}
for the Hernquist model;
\begin{equation}
\label{A-bur}
    A_\text{BUR}(r)\approx \left(1-\frac{2M_\text{BH}}{r}\right)\bigg(1-\frac{2 M r}{a_0^2\ln[a_0^2/(a_0^2+r_c^2)]}\bigg),
\end{equation}

\begin{equation}
\label{m-bur}
    m_\text{BUR}(r)\approx M_\text{BH}-\frac{M(r-2M_\text{BH})^2}{a_0^2\ln[a_0^2/(a_0^2+r_c^2)]},
\end{equation}

\begin{equation}
\label{rho-bur}
    \rho_\text{BUR}(r)\approx \frac{M(2M_\text{BH}-r)}{2\pi a_0^2r^2\ln[a_0^2/(a_0^2+r_c^2)]},
\end{equation}
for the Burkert model; and
\begin{equation}
\label{A-nfw}
    A_\text{NFW}(r)\approx \left(1-\frac{2M_\text{BH}}{r}\right)\bigg(1-\frac{M\,r\,(a_0+r_c)}{a_0^2\,[r_c-(a_0+r_c)\ln{(1+r_c/a_0)}]}\bigg),
\end{equation}

\begin{equation}
\label{m-nfw}
    m_\text{NFW}\approx  M_\text{BH}-\frac{M(a_0+r_c)(r-2M_\text{BH})^2}{2a_0^2[r_c-(a_0+r_c)\ln(1+r_c/a_0)]},
\end{equation}

\begin{equation}
\label{rho-NFW}
    \rho_\text{NFW}(r)\approx \frac{M(r-2M_\text{BH})(r_c+a_0)}{4\pi a_0^2\,r^2\left(-r_c+(a_0+r_c)\ln[1+r_c/a_0]\right)},
\end{equation}
for the NFW model; 
where the subscripts ``BUR", ``NFW" and ``HQ" represent Burkert, NFW and Hernquist, respectively.
From Eqs. \eqref{A-bur}, \eqref{m-bur}, \eqref{A-nfw}, \eqref{m-nfw}, \eqref{A-hq} and \eqref{m-hq}, 
we see that when $M=0$, the spacetimes reduce to the Schwarzschild one.
To the leading order of $M/a_0$,
the density profiles $\rho(r)\sim M/(r a_0^2)$,
result in a constant 
gravitational force acting on the SCO due to DM halos $\sim \rho(r)r^3/r^2\sim M/a_0^2$.

Combining Eqs. \eqref{A-bur}-\eqref{rho-hq}, \eqref{orbital3} and \eqref{period}, and expanding about $R_s/p$ to the second order, we get the orbital period 
\begin{equation}
\label{HQP}
\begin{split}
P_\text{HQ}\approx &P_\text{GR}+\frac{2\pi p^{3/2}}{\sqrt{M_\text{BH}}(1-e^2)^{3/2}} \left[\frac{M}{a_0}+\frac{3R_\text{s}(1-e^2)}{2p}\frac{M}{a_0}+\frac{5M^2}{6a_0^2}\right.\\
&\left.\qquad \qquad -\frac{3M}{a_0^2(1-e^2)^2}\frac{p^2}{R_s}-\frac{M p(11-2e^2)}{2a_0^2(1-e^2)}\right]
\end{split}
\end{equation}

\begin{equation}
\label{BURP}
\begin{split}
P_\text{BUR}\approx &P_\text{GR}-\frac{2\pi p^{3/2}}{\sqrt{M_\text{BH}}(1-e^2)^{3/2}} \left[\frac{3M}{a_0^2(1-e^2)^2 \log[ (a_0^2+r_c^2)/a_0^2]}\frac{p^2}{R_s}\right.\\
&\left.\qquad\qquad +\frac{M p(11-2e^2)}{2a_0^2(1-e^2)\log[(a_0^2+r_c^2)/a_0^2]}\right]
\end{split}
\end{equation}

\begin{equation}
\label{NFWP}
\begin{split}
P_\text{NFW}\approx &P_\text{GR}-\frac{2\pi p^{3/2}}{\sqrt{M_\text{BH}}(1-e^2)^{3/2}} \left[\frac{M p(11-13e^2+2e^4+6p/R_\text{s})(a_0+r_c)}{4a_0^2(1-e^2)^2(-r_c+(a_0+r_c)\log[1+r_c/a_0])}\right],
\end{split}
\end{equation}
where the orbital period without DM halos is
\begin{equation}
\label{pgrres1}    
P_\text{GR}=\frac{2\pi p^{3/2}}{\sqrt{M_\text{BH}}(1-e^2)^{3/2}} \left[1+\frac{3R_\text{s}(1-e^2)}{2p}+\frac{3R_\text{s}^2(1-e^2)(4+5\sqrt{1-e^2})}{8p^2}\right].
\end{equation}
Combining Eqs. \eqref{A-bur}-\eqref{rho-hq}, \eqref{orbital3} and \eqref{phi}, and expanding about $R_s/p$ to the second order, we get the orbital precession,
\begin{equation}
\label{HQPhi}
\Delta\phi_\text{HQ}\approx \Delta\phi_\text{GR}-\frac{6\pi M p(1+e^2+3\sqrt{1-e^2}+2p/R_\text{s})}{a_0^2(1-e^2)^{3/2}},
\end{equation}

\begin{equation}
\label{BURPhi}
\Delta\phi_\text{BUR}\approx \Delta\phi_\text{GR}-\frac{2\pi M p(1+e^2+3\sqrt{1-e^2}+2p/R_\text{s})}{a_0^2(1-e^2)^{3/2}\log[(a_0^2+r_c^2)/a_0^2]},
\end{equation}

\begin{equation}
\label{NFWPhi}
\Delta\phi_\text{NFW}\approx \Delta\phi_\text{GR}-\frac{\pi Mp(5-3e^2+2\sqrt{1-e^2}+2(1+\sqrt{1-e^2})p/R_\text{s}) (a_0+r_c)}{a_0^2(1-e^2)^{3/2}(-r_c+(a_0+r_c)\log[(a_0+r_c)/a_0])},
\end{equation}
where the orbital precession without DM halos is
\begin{equation}
\label{phigrres1}
\Delta\phi_\text{GR}=\frac{3\pi R_s}{p}+\frac{3\pi(18+e^2)R_\text{s}^2}{8p^2}.
\end{equation}


%

\end{document}